\documentclass[10pt,journal,compsoc]{IEEEtran}
\usepackage{algorithm}
\usepackage{algpseudocode}

\usepackage{amsmath}
\newcommand{\algoref}[1]{Algorithm~\ref{#1}}

\usepackage{hyperref}

\usepackage{graphics}
\usepackage{url}
\usepackage[nocompress]{cite}
\usepackage{amsmath,amssymb,amsfonts}

\usepackage{graphicx}
\usepackage{textcomp}
\usepackage{xcolor}
\usepackage{booktabs}
\usepackage{subcaption}
\usepackage{cleveref} 
\usepackage{soul}
\usepackage{multirow}
\usepackage{indentfirst}    
\usepackage{array}      
\usepackage{ragged2e}
\usepackage{threeparttable}     
\usepackage{color}

\ifCLASSOPTIONcompsoc
  \usepackage[nocompress]{cite}
\else
  \usepackage{cite}
\fi
\ifCLASSINFOpdf
\else
\fi
\begin{document}
%
\title{Identifying Root Cause of bugs by Capturing Changed Code Lines with Relational Graph Neural Networks}
%
%
%
%

\author{Jiaqi~Zhang,
        Shikai~Guo,
        Hui~Li,
        Chenchen~Li,   
        Yu~Chai,
        Rong~Chen
        
\IEEEcompsocitemizethanks{

\IEEEcompsocthanksitem J. Zhang, S. Guo are with the School of Information Science and Technology, Dalian Maritime University, Dalian, China, and the Dalian Key Laboratory of Artificial Intelligence, Dalian, China. E-mail: zhangjiaqi@dlmu.edu.cn, shikai.guo@dlmu.edu.cn

\IEEEcompsocthanksitem H Li and R Chen is with the School of Information Science and Technology, Dalian Maritime University, Dalian, China. E-mail: li\_hui@dlmu.edu.cn, rchen@dlmu.edu.cn

\IEEEcompsocthanksitem C. Li is with the School of Information Science and Technology, Liaoning Normal University, Dalian, China. E-mail: lcc@lnnu.edu.cn

\IEEEcompsocthanksitem Y Chai is with the School of Oceanography, Shanghai Jiao Tong University, Shanghai, China. E-mail: cy7372@sjtu.edu.cn

}
\thanks{Manuscript received October 7, 2024; revised October 7, 2024.}}

%
%

\markboth{IEEE Transactions on Consumer Electronics}%
{Shell \MakeLowercase{\textit{Zhang et al.}}: Bare Demo of IEEEtran.cls for Computer Society Journals}
%



\IEEEtitleabstractindextext{%
\begin{abstract}
\justifying{
The Just-In-Time defect prediction model helps development teams improve software quality and efficiency by assessing whether code changes submitted by developers are likely to introduce defects in real-time, allowing timely identification of potential issues during the commit stage. 
However, two main challenges exist in current work due to the reality that all deleted and added lines in bug-fixing commits may be related to the root cause of the introduced bug: 1) lack of effective integration of heterogeneous graph information, and 2) lack of semantic relationships between changed code lines. 
To address these challenges, we propose a method called RC-Detection, which utilizes relational graph convolutional network to capture the semantic relationships between changed code lines. 
RC-Detection is used to detect root-cause deletion lines in changed code lines, thereby identifying the root cause of introduced bugs in bug-fixing commits. 
Specifically, the RC-Detection consists of three components: the graph construction component, the graph type conversion component, and the root cause detection component. 
The graph construction component analyzes the source code of bug-fixing commits to construct a heterogeneous graph representation by extracting added/deleted nodes based on added/deleted lines and extracting edges according to the relationships between the nodes. 
Next, to address the challenge of varying feature dimensions and the difficulty of integrating information in the heterogeneous graph of changed code lines, the graph type conversion component merges different types of nodes/edges into a unified set of nodes/edges and the type information of each node/edge is encoded as an additional vector. 
This process unifies the heterogeneous graph data into homogeneous graph data while preserving the type characteristics of different nodes and edges, thereby facilitating the integration of information from various nodes and edges. 
Finally, the root cause detection module uses a node embedding layer to obtain embedding vectors for the corresponding code statements, followed by a relational graph convolutional layer to capture the semantic relationships between the changed code lines and generate prediction labels. 
Ultimately, the root cause deletion lines in the bug-fixing commit are identified through a ranking layer applied to the deleted nodes. To evaluate the effectiveness of RC-Detection, we used three datasets that contain high-quality bug-fixing and bug-introducing commits.
Extensive experiments were conducted to evaluate the performance of our model by collecting data from 87 open-source projects, including 675 bug-fix commits.
The experimental results show that, compared to the most advanced root cause detection methods, RC-Detection improved Recall@1, Recall@2, Recall@3, and MFR by at 4.107\%, 5.113\%, 4.289\%, and 24.536\%, respectively.  
}
\end{abstract}

}

\maketitle

\IEEEdisplaynontitleabstractindextext

%
\IEEEpeerreviewmaketitle

\IEEEraisesectionheading{\section{Introduction}\label{sec:introduction}}

%
%
%
%
\IEEEPARstart{I}{n} modern software development, ensuring software quality and efficiency is a critical goal for all development teams. 
As project size and complexity increase, manually detecting and fixing defects in the code becomes increasingly difficult and time-consuming\cite{Minelli_Mocci_Lanza_2015}, \cite{Branch_Jackson_Laviolette_Frankel_1986}. 
To address this challenge, Just-In-Time (JIT) defect prediction models have emerged\cite{Kamei_Shihab_Adams_Hassan_Mockus_Sinha_Ubayashi_2013}, \cite{10388006}, \cite{10286288}, \cite{zhang2023anomaly}, \cite{Mockus_Weiss_2002}. 
These models help development teams identify potential issues at the commit stage by assessing in real-time whether code changes submitted by developers are likely to introduce defects\cite{Kim_Whitehead_Zhang_2008}. 
However, the effectiveness of JIT defect prediction models heavily relies on accurately labeling the code changes that introduce bugs\cite{Fan_Xia_da_Costa_Lo_Hassan_Li_2021}. 
If the model fails to accurately label which code changes introduce defects, it faces the problem of high false positive and false negative rates, ultimately affecting its overall performance. 
Therefore, accurately identifying and labeling these defect-inducing code changes is crucial\cite{da_Costa_McIntosh_Shang_Kulesza_Coelho_Hassan_2017}, \cite{Kim_Zimmermann_Pan_Jr_Whitehead_2006}, not only to improve the performance of defect prediction models but also to provide valuable feedback to development teams, helping them better understand the root causes of defects and improve coding practices\cite{Kamei_Shihab_Adams_Hassan_Mockus_Sinha_Ubayashi_2013}, \cite{Kamei_Matsumoto_Monden_Matsumoto_Adams_Hassan_2010}.

The SZZ algorithm, as a widely used defect identification technique, plays an important role in identifying and labeling whether code change commits introduce defects. 
Sliwerski et al. proposed the original SZZ algorithm (B-SZZ)\cite{sliwerski2005changes}, which traces back to the last commit that was changed in the bug-fixing commit and marks it as a bug-inducing commit. 
However, due to noise present in bug-fixing commits, the accuracy of the B-SZZ algorithm is relatively low. 
To address this issue, many existing SZZ algorithms and their variants improve the accuracy of the SZZ algorithm by using static methods to filter out noise. 
Kim et al. introduced AG-SZZ\cite{Kim_Zimmermann_Pan_Jr_Whitehead_2006}, which improves B-SZZ\cite{sliwerski2005changes} by using annotation graphs to filter out blank lines, marked comments, and meaningless changes in bug-fixing commits. 
Da Costa et al. proposed MA-SZZ\cite{da_Costa_McIntosh_Shang_Kulesza_Coelho_Hassan_2017}, which filters out meta-changes that do not alter the source code, such as branch changes, merge changes, and attribute changes, from potential bug-inducing changes. 
These methods attempt to filter out unnecessary changes in bug-fixing commits, such as comments and refactoring operations. 
In addition, Neto et al. introduced the RA-SZZ algorithm\cite{Neto_da_Costa_Kulesza_2018}, which integrates the RefDiff tool, capable of detecting 13 types of refactoring operations, but it may still contain unrelated lines. 
Therefore, Tang et al. proposed NEURAL-SZZ\cite{tang2023neural}, a deep learning-based method that builds a heterogeneous graph attention network model. 
RC-Detection captures the semantic relationships between each deleted line and other deleted and added lines to detect the root-cause deletion lines in bug-fixing commits. Despite the progress made in these studies on SZZ, there are still two challenges.

\textbf{Challenge 1: Lack of Effective Integration of Heterogeneous Graph Information.} 
In bug-fixing commits, the changed code lines typically include deleted lines and added lines, and various types of relationships exist among all these statements. 
To analyze these relationships, changed code lines are often structured into heterogeneous graphs. However, in bug-fixing commits, there is often a frequent alternation between added and deleted lines. Meanwhile, there are complex relationships among changed code lines, including control dependencies, data dependencies, and call relationships. Consequently, the nodes and edges in the heterogeneous graph structure of changed code lines frequently exhibit different dimensional characteristics. 
This makes it challenging for models to effectively integrate information from heterogeneous graphs\cite{Sun_Han_2013}, thereby limiting RC-Detection’s learning capability. 
For example, Tang et al. proposed the NEURAL-SZZ method \cite{tang2023neural}, which similarly extracts bug-fixing commit data into a heterogeneous graph. 
Although they attempted to build a heterogeneous graph attention network model to address this issue, it still tends to result in model over-parameterization. 
Therefore, how to integrate and fully utilize the information from the varying dimensional features in a heterogeneous graph remains a key challenge.

\textbf{Challenge 2: Lack of Semantic Relationships Between Changed Code Lines.}
In bug-fixing commits, the semantic relationships between the changed code lines are not fully utilized, leading to significant interference from noise in the existing SZZ algorithm. 
Existing research primarily focuses on filtering out unnecessary or meaningless changes in bug-fixing commits, such as comments and refactoring operations. 
However, since there are both deleted and added lines in changed code, both types of code lines may be related to the lines that introduced the actual bug. 
Therefore, whether the semantic relationships between the changed code lines are fully captured is crucial for identifying whether the code change commit introduces defects. 
If we only focus on filtering out unnecessary or meaningless changes in bug-fixing commits while ignoring these relationships, valuable defect identification information could be lost. 
Therefore, how to capture and utilize the semantic relationships between different changed code lines has become a significant challenge.

To address these challenges, we propose a method called RC-Detection, which uses a relational graph convolutional network \cite{Schlichtkrull_Kipf_Bloem_van·den_Berg_Titov_Welling_2018} to capture the semantic relationships between changed code lines, aiming to detect the root-cause deletion lines in code change commits. 
RC-Detection is primarily composed of the following three components: the graph construction component, the graph type conversion component, and the root cause detection component. 
Given a bug-fixing commit, the graph construction component analyzes the changed code lines, extracting added and deleted lines as nodes, and extracting edges based on the relationships between code lines, thereby constructing the graph data. 
The graph type conversion component merges different types of nodes/edges into a unified set of nodes/edges and the type information of each node/edge is encoded as an additional vector. This process unifies the heterogeneous graph data into homogeneous graph data while preserving the type characteristics of different nodes and edges. Thus, the graph type conversion component facilitates the integration of information from various nodes and edges, addressing \textbf{Challenge 1}. 
The root cause detection component will vectorize the graph data obtained from the previous component and then pass it into the relational graph convolutional layer to capture the semantic relationships between the changed code lines. 
RC-Detection predicts a probability label for each deleted code line indicating whether it is the root cause. 
Finally, the deletion line nodes are ranked with a pairwise based ranking method according to the predicted labels, thereby identifying the root-cause deletion lines of bugs in the bug-fixing commit, addressing \textbf{Challenge 2}.

Extensive experiments were conducted to evaluate the performance of our model by collecting data from 87 open-source projects\cite{Wen_Wu_Liu_Tian_Xie_Cheung_Su_2019}, \cite{Song_Lin_Ng_Wu_Peng_Dong_Mei_2021}, \cite{Neto_Costa_Kulesza_2019}, including 675 bug-fix commits. The experimental results show that compared to state-of-the-art root cause detection methods for bugs, RC-Detection improved Recall@1, Recall@2, Recall@3, and MFR by at least 4.107\%, 5.113\%, 4.289\%, and 24.536\%, respectively. 
These results demonstrate the effectiveness of RC-Detection in detecting the root causes of bugs. 


\par The main contributions of this paper can be summarized as follows: 

\begin{itemize}
    \item We propose RC-Detection, which uses the relational graph convolutional network to capture the semantic relationships between changed code lines, aiming to detect root-cause deletion lines in code change commits.
    
    \item We conducted a series of experiments to verify the effectiveness of the RC-Detection method. RC-Detection achieved scores of 0.811, 0.884, 0.924, and 1.830 on Recall@1, Recall@2, Recall@3, and MFR, respectively, which mark improvements of 4.107\%, 5.113\%, 4.289\%, and 24.536\%b over the state-of-the-art methods.
    
    \item We intend to release the complete code and evaluation datasets of RC-Detection to foster reproducibility and facilitate its utilization by further investigations.\href{https://github.com/Leling666/RC-Detection}{(https://github.com/Leling666/RC-Detection)}. 
\end{itemize}

The rest of this paper is organized as follows. Our motivation is discussed in Section 2. 
Section 3 introduces the main components of RC-Detection. Experimental setup and results are presented in Sections 4 and 5, respectively. 
In Section 6, we discuss the threats to validity in detail. 
Section 7 presents an overview of related work in the field Finally, Section 8 concludes the paper and outlines future work.

\section{BACKGROUND AND MOTIVATION}

\begin{figure}[!t]
	\centering
	\includegraphics[width=\linewidth]{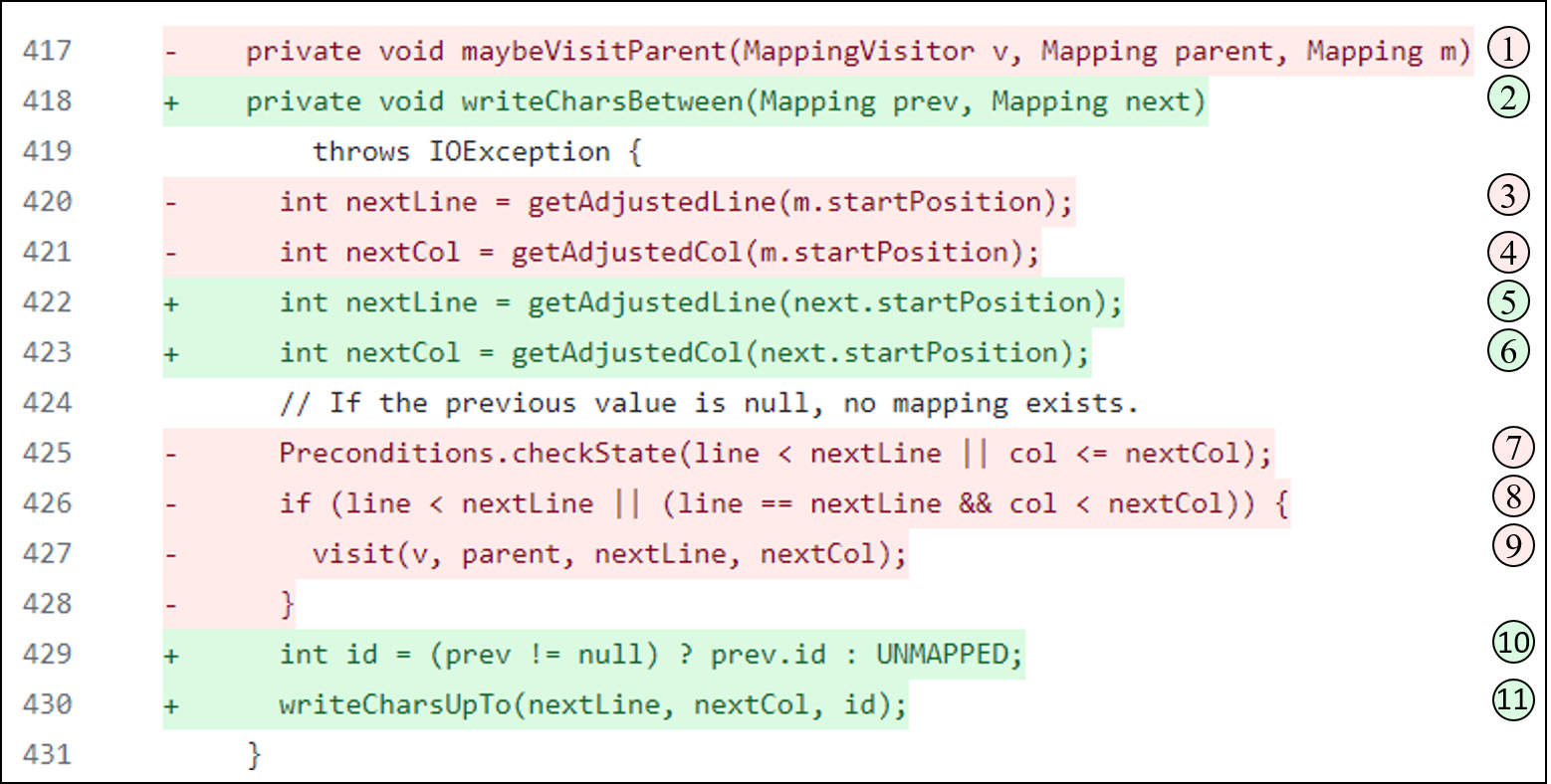}
	\caption{A motivation example of changed code lines}
    \label{fig:Example}
\end{figure}

\figurename~\ref{fig:Example} illustrates an example of a bug-fixing commit in the project of closure-compiler. 
This example snippet includes seven deleted lines and five added lines from the file “\emph{PeepholeFoldConstants.java.}” 
Based on the commit information, the root cause of the bug corresponds to line 427 in this file. Running the original SZZ algorithm directly on all the modifications in this bug-fixing commit in this scenario could introduce a significant amount of noise because not all the changes are the factors of the bug, thus impacting the accuracy of the SZZ algorithm. 
To address this issue, the root-cause code lines within the commit need to be identified, so that other noise can be eliminated and only the root-cause lines are used as input to the SZZ algorithm.
Therefore, it is crucial to identify the root-cause code lines. We observe that the relationships between changed lines in a bug-fixing commit can help in identifying the root cause. 
In the motivation example, if we only consider the deletion of line 427, it may not be clear what caused the bug. 
However, considering the relationships between line 427 and other lines, such as line 407, 417, 420, and 421, provides more insights. 
For instance, the data flow relationships between lines 427 and 417, 420, and 421. Previous research, however, has primarily focused on filtering out unnecessary or meaningless changes in bug-fixing commits, such as comments and refactoring operations. As a result, the relationships between changed code lines have not been fully utilized. This leads to Challenge 2, lack of semantic relationships between changed code lines. 

\par Additionally, we found that in Figure 1, added and deleted lines frequently alternate, and there are complex relationships such as control, data dependencies, and calls between the changed code lines. For instance, the values of deleted line 3 and 4 depend on the input m from deleted line 1, while the values of deleted line 3 and 4 influence deleted line 7, 8, and 9. Furthermore, there are sequential control flow relationships among deleted line 7, 8, and 9. These complex relationships lead to a heterogeneous graph structure extracted from the changed code lines, where nodes and edges often have different dimensional characteristics, making it difficult to integrate effective information. This situation can easily cause model overparameterization, which significantly affects the model's learning ability. This motivates the emergence of Challenge 1, the lack of effective integration of heterogeneous graph information. 

\par Considering these factors, we propose the RC-Detection method, which uses the RGCN to capture the semantic relationships between changed code lines, aiming to detect the root-cause deletion lines in code change commits.

\section{RC-Detection Model}

In this section, 3.1 provides an overview of the model framework. 
Next, we describe the graph construction process in 3.2. 
Following that, the methods for the graph type conversion component are introduced in 3.3. Finally, we explain the implementation of the root cause detection component in section 3.4. 

\begin{figure*}[!t]
	\centering
	\includegraphics[width=18cm]{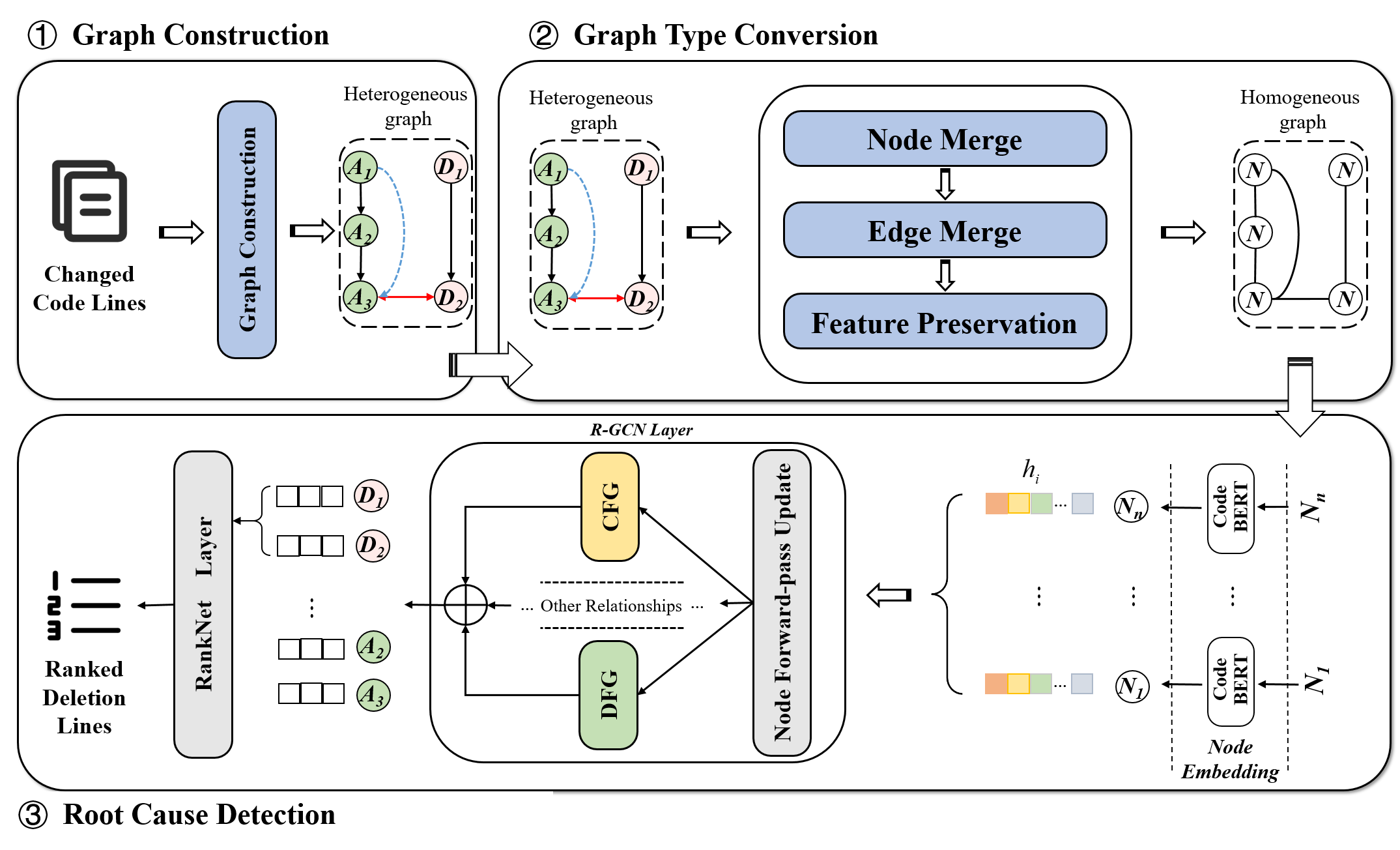}
	\caption{The overall framework of RC-Detection}\label{fig:framework}
\end{figure*}

\subsection{Overview}
To address these challenges, we propose a method called RC-Detection, which uses RGCN to capture the semantic relationships between changed code lines, aiming to detect the root-cause deletion lines in code change commits. The framework of the RC-Detection method is shown in \figurename~\ref{fig:framework}. The RC-Detection method is primarily composed of three components: the graph construction component, the graph type conversion component, and the root cause detection component. 

\par Firstly, In the graph construction component, RC-Detection first process the data of the changed code lines. Given a bug-fixing commit, graph construction component analyze the modified code lines, extracting added and deleted lines as nodes, while also extracting edges based on the relationships between the code lines, thereby constructing heterogeneous graph data. 

\par Next, the data is passed to the graph type conversion component. Here, the graph type conversion component merge different types of nodes/edges into a unified set of nodes/edges, with the type information of each node/edge encoded as an additional vector. The graph type conversion component simplifies the graph's structure by converting the heterogeneous graph data into homogeneous graph data while preserving the type features of different nodes and edges, facilitating the integration of information from various nodes and edges, thus addressing Challenge 1. 

\par Finally, in the root cause detection component, the graph data obtained from the previous module is first vectorized using CodeBERT, and then passed through RGCN to capture the semantic relationships between the changed code lines. Each deleted code line is assigned a probability label indicating whether it is the root cause. 
The deleted lines are then ranked based on these probability labels using a pairwise based ranking method. 
In this way, RC-Detection detect the deleted lines most likely to be the root cause of the bug in the bug-fixing commit, thereby addressing Challenge 2.

\subsection{Graph Construction Component}
\par
\begin{figure}[!t]
	\centering
	\includegraphics[width=\linewidth]{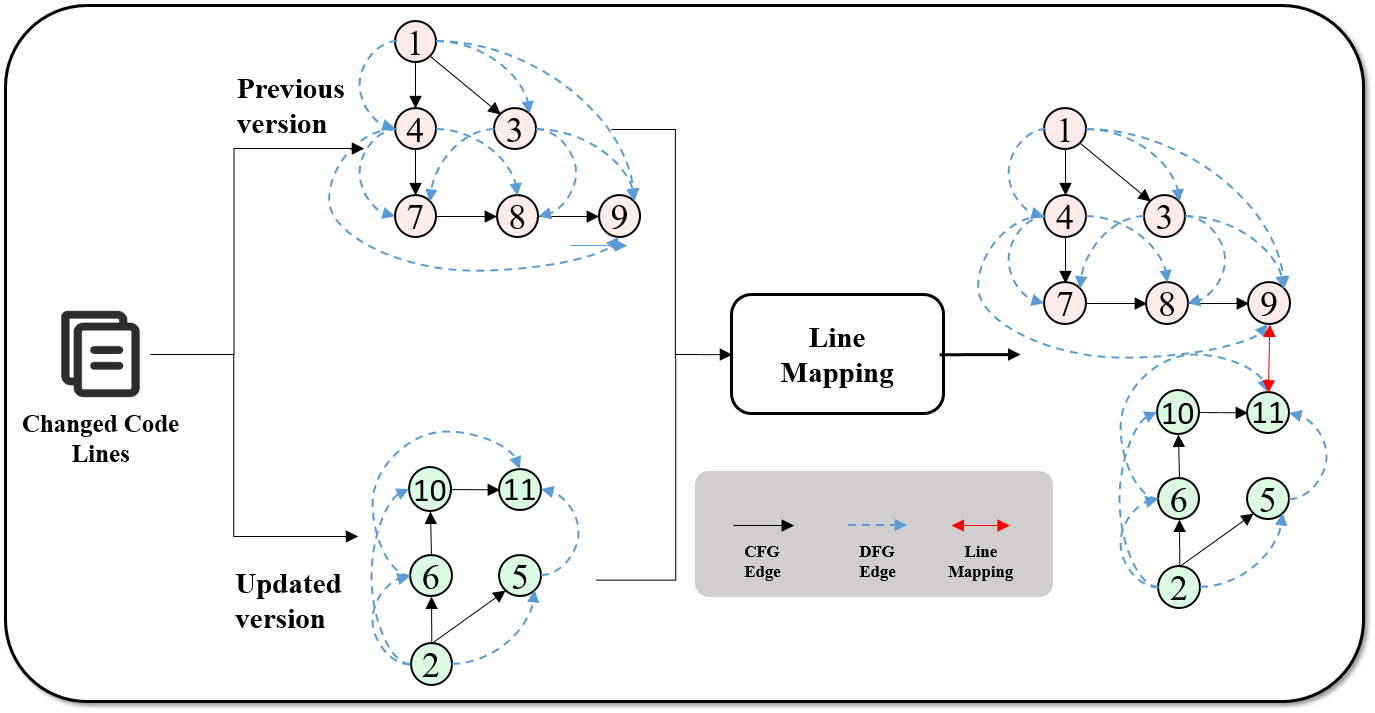}
	\caption{A practical example of graph construction}
     \label{fig:GC}
\end{figure}

 In this work, the graph construction component represent and learn the hierarchical structure of the code by extracting the graph structure of the changed code lines. 
\figurename~\ref{fig:GC} illustrates the graph construction process.

\par In bug-fixing commits, the changed code lines typically consist of two types of statements: deleted lines and added lines, and there are different types of relationships between these lines.
To represent the two types of code lines and distinguish between their various relationships, the graph construction component extract deleted and added lines from the bug-fixing commit as nodes and construct edges based on their relationships, ultimately forming a heterogeneous graph.

\begin{algorithm}
\caption{Procedure for Graph Construction}
\label{algorithm1}
\begin{algorithmic}[1]

    \Require sourceCodeOld: $CO$, sourceCodeNew: $CN$
    \Ensure The Graph of Changed Code Lines: $G$
    \For{$delLine$ $\in$ $CO$}
        \State $delNodes$ $\leftarrow$ $delNodes$ + node($delLine$)
    \EndFor
    \State $AST_{pre}$ = constructAST($delNodes$)
    \For{$addLine$ $\in$ $CN$}
        \State $addNodes$ $\leftarrow$ $addNodes$ + node($addLine$)
    \EndFor
    \State $AST_{new}$ = constructAST($addNodes$)
    \For{$relation$ $\in$ $allRelations$}
        \State $relPreList$ $\leftarrow$ $relPreList$ + relGraph($delNodes$)
        \State $relNewList$ $\leftarrow$ $relNewList$ + relGraph($addNodes$)
    \EndFor
    \For{$graph, relation$ $\in$ $relPreList$}
        \State ASTpre.addEdges = explorePaths($graph$, $relation$)
    \EndFor
    \For{$graph$ $\in$ $relNewList$}
        \State ASTnew.addEdges = explorePaths($graph$, $relation$)
    \EndFor
    \For{$Dnode$ $\in$ $delNodes$}
        \For{$Anode$ $\in$ $addNodes$}
            \If{Mapping($Dnode$, $Anode$)}
            \State \text{addMapEdge($AST_{pre}$, $AST_{new}$, $Dnode$, $Anode$)}
            \EndIf
        \EndFor
    \EndFor
    \State \Return $G$ = combineGraph($AST_{pre}$, $AST_{new}$)

\end{algorithmic}
\end{algorithm}

\par As shown in \algoref{algorithm1}, for a bug-fixing commit, the graph construction component first analyze the source code of both the previous and updated versions, constructing abstract syntax trees ($AST_{pre}$ and $AST_{new}$) for each version. 
The graph construction component map the deleted lines to $AST_{pre}$, marking them as deletion nodes, and similarly map the added lines to $AST_{new}$, marking them as addition nodes, thus extracting the deletion and addition nodes (lines 1-8). 
Next, based on these extracted nodes, the graph construction component construct edges by considering the different relationships between them. 
Here, RC-Detection utilize Control Flow Graphs (CFG)\cite{Allen_1970}, Data Dependency Graphs (DDG)\cite{Ferrante_Ottenstein_Warren_1987}, Call Graphs (CG)\cite{Ryder_1979}, and Class Member Reference Graphs (CMFG) to represent the relationships between nodes. 
We  construct these graphs separately for the source code of both versions (lines 9-12) and then use depth-first search (DFS) to explore paths between nodes in each graph. If a path exists between two nodes, the graph construction component add an edge between them and label it according to the graph type (lines 13-18). 
This process results in two graphs: one containing only deletion nodes derived from the previous version of the source code, and another containing only addition nodes derived from the updated version. 
Finally, if a mapping relationship exists between a deleted line and an added line, the graph construction component add a line-mapping edge between the corresponding nodes in both graphs (lines 19-25). 
By connecting the two graphs through line mapping edges, the graph construction component obtain the final graph of the changed code lines (line 26).

\figurename~\ref{fig:GC} illustrates the process of graph construction based on the motivational example (shown in \figurename~\ref{fig:Example}).The node IDs in this graph correspond to the IDs in \figurename~\ref{fig:Example}. First, RC-Detection extract nodes from the changed lines in the bug-fixing commits. For instance, the deleted line 417 corresponds to the deleted node 1 in $AST_{pre}$, while the added line 2 corresponds to the added node 2 in $AST_{new}$.Next, RC-Detection use a depth-first search algorithm to extract edges from the various relationship graphs between the nodes.For example, RC-Detection determine that deleted node 1 can establish a path to added deleted 3 without going through any other added nodes in the CFG. Therefore, RC-Detection add a control flow edge between them. Similarly, RC-Detection determine that added node 2 can establish a path to added node 5 without going through any other added nodes in the DDG. Therefore, RC-Detection add a data dependency edge between them.RC-Detection repeat this process for each type of graph to retrieve all edges between any pair of related nodes. Finally, RC-Detection identify that deleted node 9 can be mapped to added node 11, so RC-Detection add a line mapping edge between them to obtain the final graph.

\subsection{Graph Type Conversion Component}

\par
\begin{figure}[!t]
	\centering
	\includegraphics[width=\linewidth]{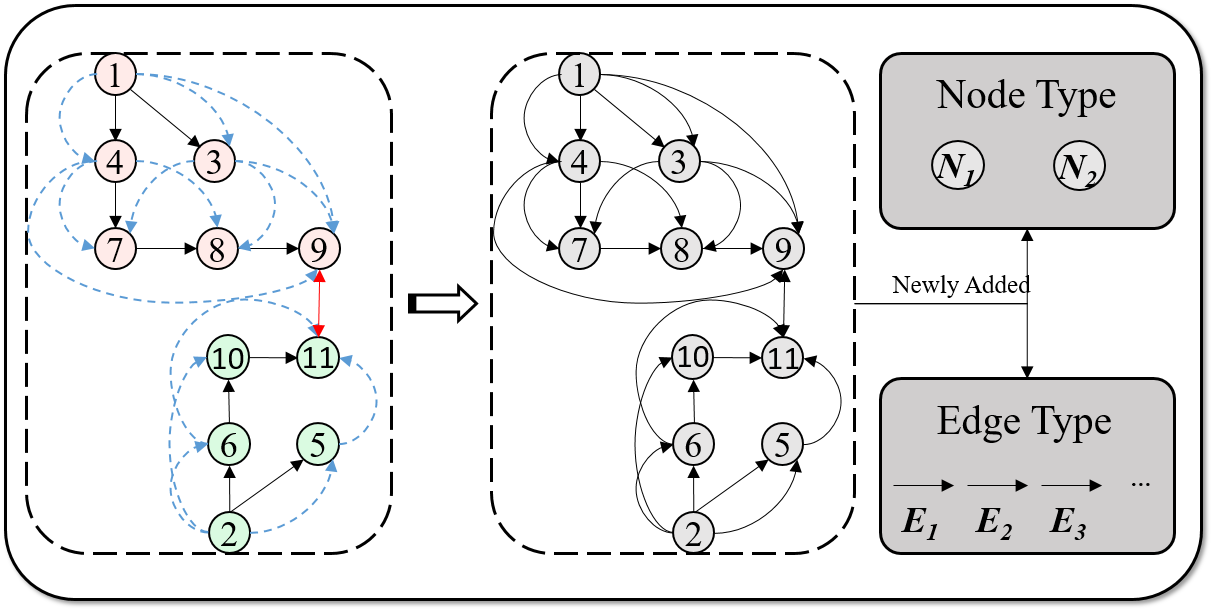}
	\caption{A practical example of graph type conversion}
    \label{fig:GTC}
\end{figure}

 After the graph construction is completed, to better integrate and utilize the information of different dimensional features in the heterogeneous graph, it is necessary to transform the heterogeneous graph data into homogeneous graph data in the graph type conversion component. \figurename~\ref{fig:GTC} illustrates our example of converting a heterogeneous graph into a homogeneous graph. In this section, we propose a method to convert the heterogeneous graph $G_H$ into a homogeneous graph $G_{Ho}$ by merging different nodes and edges into a unified set of nodes and edges, with the type information of each edge encoded as an additional vector. This way, the heterogeneous graph data is uniformly processed into homogeneous graph data while preserving the type feature information of different nodes and edges, facilitating the subsequent integration of information from different nodes and edges. The detailed process of graph type conversion is described below. 

The heterogeneous graph $G_H$ is defined as follows:
\begin{equation}
G_H = (V_H, E_H, \{X_v\}_{v \in V_H}, \{X_e\}_{e \in E_H})
\end{equation}
where $V_H$  represents the set of nodes, $E_H$ denotes the set of edges, and ${X_v}$ and ${X_e}$ correspond to the feature matrices related to nodes and edges, respectively. 

First, to ensure the compatibility of features among different types, the graph type conversion component traverse the storage of nodes and edges to obtain the dimensions of each feature:
\begin{equation}
\text{size}_{key} = \{ \text{size}(X_v) | v \in V_H, \text{for each } key \in store \}
\end{equation}
where $\text{size}_{key}$is a collection of dimensions for specific features, while $store$ is the storage structure that contains the features of nodes or edges. 

However, in certain cases, features may be missing across different types. To maintain consistency, the graph type conversion component fills in these missing features with dummy values. If a missing feature value is detected, the graph type conversion component will fill in the missing features according to the following rules: 
\begin{equation}
X_dummy = \begin{cases} 
\text{NaN} & \text{if } type(X) = \text{  float} \\
\text{False} & \text{if } type(X) = \text{  boolean} \\
-1 & \text{if } type(X) = \text{  integer}
\end{cases}
\end{equation}
where $X_{dummy}$ is the filled dummy value. $type(X)$ indicates the data type of feature $X$. $NaN$ represents a missing floating-point value, $False$ indicates a missing boolean value, and ${-1}$ signifies a missing integer value. 

Next, the graph type conversion component check the existence of each feature across all relevant node or edge types. If a feature is missing in certain types, it does not meet the consistency requirements: 
\begin{equation}
\text{valid}(key) = \text{len}(\{ \text{exists}(X_v) | v \in V_H \}) = \text{len}(store)
\end{equation}
where $key$ is the features in the $store$.
\par Finally, RC-Detection retain only those features that consistently exist across all node or edge types for further processing: 
\begin{equation}
keys \&= \{key \mid \text{valid}(key) \}
\end{equation}
where $keys$ is the set of features that consistently exist across all node or edge types in the $store$.This set will be used for subsequent operations of merging node and edge features, ensuring that the final constructed homogeneous graph is reliable and consistent. 

\par After the above steps, the graph type conversion component will merge the feature matrices of nodes and edges to form a unified feature representation. The merging of node features can be expressed as: 
\begin{equation}
X_{Ho} = \text{cat}\left(\{X_v | v \in V_H, \text{for each } key\}\right)
\end{equation}
where $X_{Ho}$ represents a unified feature matrix for the nodes in the homogeneous graph.
\par Similarly, RC-Detection merge the edge features:
\begin{equation}
Y_{Ho} = \text{cat}\left(\{Y_e | e \in E_H, \text{for each } key\}\right)
\end{equation} 
where $Y_{Ho}$ represents a unified feature matrix for the edges in the homogeneous graph.

\par During the process of merging the feature matrices, to retain the type feature information of different nodes and edges, the type information for each node and each edge will be encoded as an additional vector, referred to as $node\_type$ and $edge\_type$, respectively: 
\begin{equation}
node\_type = \text{repeat}\left(\{ \text{size} \text{ for each node type}\}\right)
\end{equation}
\begin{equation}
edge\_type = \text{repeat}\left(\{ \text{size} \text{ for each edge type}\}\right)
\end{equation}
where $node\_type$ and the $edge\_type$ represents type information for each node and each edge.To compute $node\_type$, the graph type conversion component first determine the count of each node type and then generate a repeated vector based on the quantity of nodes for each type. For example, if a certain node type has a count of $n$, its identifier $a$ will be repeated $n$ times. Thus, the final $node\_type$ vector contains the type identifiers corresponding to each node. The method for calculating $edge\_type$ is the same as that for $node\_type$.

\par Finally, the graph type conversion component encapsulate all the transformed features into a new homogeneous graph set:
\begin{equation}
G_{Ho} = (V_{Ho}, E_{Ho}, {X_{Ho}}, {Y_{Ho}}, node\_type, edge\_type)
\end{equation}
where $V_{Ho}$  represents the set of nodes, $E_{Ho}$ denotes the set of edges, and $X_{Ho}$ and $Y_{Ho}$ correspond to the feature matrices related to nodes and edges, respectively. $node\_type$ is the type information for each node and the $edge\_type$ is the type information for each edge.

\par In this way, the heterogeneous graph $G_H$  is unified into a homogeneous graph $G_{Ho}$ , while also retaining the type feature information of different nodes and edges, thus addressing Challenge 1. 

\subsection{Root Cause Detection Component}
To capture the semantic relationships between changed code lines and identify the root cause deletion lines of bus in bug\-fixing commits, RC-Detection designed the root cause detection component. 
This component consists of three parts:Node Embedding Layer,RGCN Layer and Deletion Nodes Rankling Layer.

\subsubsection{Node Embedding Layer}
\par In this layer, the root cause detection component use CodeBERT\cite{Feng_Guo_Tang_Duan_Feng_Gong_Shou_Qin_Liu_Jiang_et_al._2020} to vectorize the code line statements, embedding them into fixed-length vectors. 
CodeBERT is a pre-trained language model specifically designed for programming and natural languages, supporting various downstream tasks involving both. 
Based on the Transformer architecture\cite{vaswani2017attention}, it combines Masked Language Modeling (MLM)\cite{devlin2018bert} and Replaced Token Detection (RTD)\cite{clark2020electra} as pre-training objectives, and has been trained on a large-scale dataset of code, achieving state-of-the-art performance on many code-related tasks. 
CodeBERT captures the semantic information of the code by transforming code statements into a sequence of tokens, which are then encoded to generate vector representations for each token, making them suitable as inputs to the RGCN Layer. 
Thus, for each node, the root cause detection component use CodeBERT to obtain its corresponding embedding vector.

\subsubsection{RGCN Layer}
\par The RGCN layer follows the node embedding layer. In this layer, the root cause detection component utilize relational graph convolutional networks to capture the semantic relationships between changed code lines and predict the probability labels indicating whether each deleted code line is the root cause.

\par Relational graph convolutional  network\cite{Schlichtkrull_Kipf_Bloem_van·den_Berg_Titov_Welling_2018} is an extension of graph convolutional networks (GCN)\cite{Kipf_Welling_2016},which can operate on local graph neighborhoods\cite{Duvenaud_Maclaurin2015}, \cite{Kipf_Welling_2016} to large-scale relational data. This and related methods such as graph neural networks\cite{Scarselli_Gori_Ah_Chung2009} can be understood as special cases of a simple differentiable message-passing framework:

\begin{equation}
h^{(l+1)}_{i} = \sigma\left(\sum_{m\in M_{i}} g_{m}
\left( h^{(l)}_{i} , h^{(l)}_{j}       
\right)
\right)
\end{equation}
where $h_{i}^{(l)}\in R^{d^{(l)}}$ is the hidden state of node $v_{i}$ in the $l$-th layer of the neural network, with $d^{(l)}$ being the dimensionality of this layer's representations. Incoming messages of the form $gm(\cdot, \cdot)$ are accumulated and passed through an element-wise activation function $\sigma(\cdot)$. $M_{i}$ denotes the set of incoming messages for node vi and is often chosen to be identical to the set of incoming edges. 

The core idea of relational graph convolutional network is to introduce independent weight matrices for each type of edge, performing weighted linear transformations and aggregation of the features of neighboring nodes, thereby utilizing the information from different types of nodes and edges in the graph. Specifically, relational graph convolutional network calculates the weighted aggregation of neighboring features for each node based on the type of each edge, merging this with the node's own features to capture the semantic relationships between nodes. In this way, relational graph convolutional network can effectively capture the complex semantic relationships between changed code lines. 

\par
\begin{figure}[!t]
	\centering
	\includegraphics[width=\linewidth]{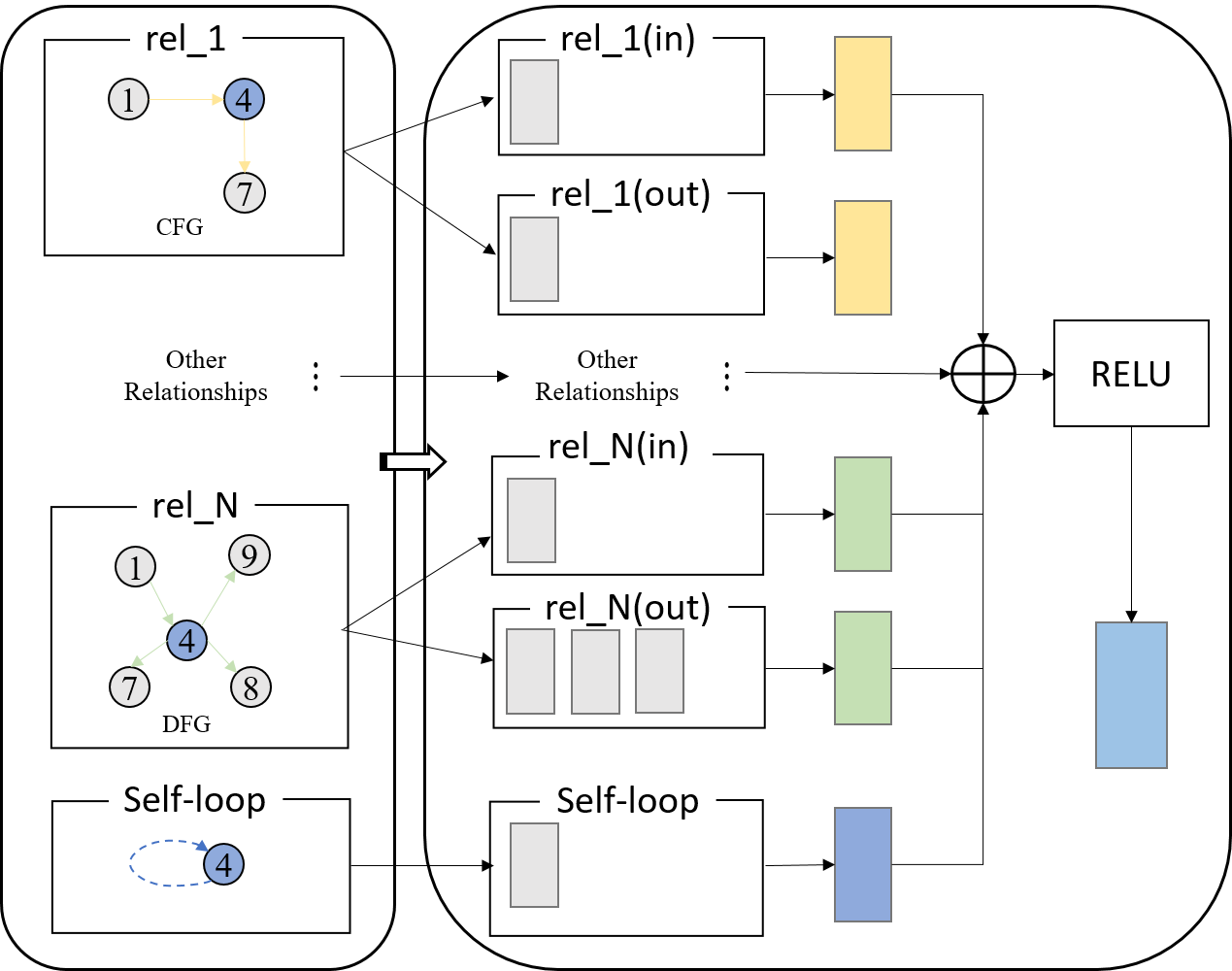}
	\caption{Diagram for computing the update of a single graph node (Blue) in the RGCN Layer.}
     \label{fig:node_update}
\end{figure}

 The overall relational graph convolutional network employs a multi-layer stacking method, where the output of the previous layer serves as the input for the next layer. The root cause detection component define the following simple propagation model to compute the forward updates of entities or nodes in a multi-relational graph: 
\begin{equation}
h^{l+1}_{i}=\sigma\left(\sum_{r\in R}\sum_{j \in N^{r}_{i}} \frac{1}{c_{i,r}} W^{(l)}_{r} h_{j}^{(l)} + W_{0}^{(l)} h_{i}^{(l)} \right)
\end{equation}
where $h_{i}^{(l)}\in R^{d^{(l)}}$is the hidden state of the node $v_{i}$ at $l$-th layer, and $d^{(l)}$is the dimension of this layer. $N^{r}_{i}$ represents the neighbor set of node $i$ under relation $r\in R$. $c_{i,r}$is a problem-specific normalization constant that can be learned or selected in advance(e.g. $c_{i,r} =\left| N^{r}_{i}\right|$ ). The two main parameters in the propagation model are the dimension transformation matrices: $W^{(l)}_{r}$ is the weight matrix associated with relation $r$ at layer$l$ , responsible for linearly transforming the features of neighbor nodes. $ W^{(l)}_{0}$ is the weight matrix for self-connections, used to merge the features of the node itself. Through these matrices, the relational graph convolutional network can effectively aggregate information from neighbor nodes across different types of relations and update the hidden state of the nodes. \figurename~\ref{fig:node_update} illustrates the computation process for updating a single node, demonstrating how to update node states by aggregating neighbor information and combining self-connection features.

\par When applying this propagation model to multi-relation data, a core issue is that as the number of relations increases, the number of model parameters\cite{glorot2010understanding} also grows rapidly. This situation can easily lead to overfitting in practice. To address this problem, relational graph convolutional network introduces two regularization methods: \emph{basis-} and \emph{block-diagonal-}decomposition. Both methods alleviate overfitting by reducing the number of parameters relational graph convolutional network needs to learn. Specifically, basis decomposition represents the weight matrices for different relations as a set of shared basis functions combined with their corresponding weights, while block diagonal decomposition structures the weight matrices into a block diagonal form, thereby reducing the number of parameters. Both methodes effectively tackle the parameter explosion problem in datasets with a large number of relations.

\par \textbf{Basis Decomposition}: Basis decomposition is an effective method for reducing the number of parameters by sharing matrix parameters, which can be seen as a way of sharing weights among different relation types. This method alleviates the overfitting problem by allowing rare relation types to share parameter updates with more frequent relation types. Specifically, in basis decomposition, each$W^{(l)}_{r}$is defined as follows:

\begin{equation}
W_r^{(l)} = \sum_{b=1}^{B} a_{rb}^{(l)} V_b^{(l)}
\end{equation}

As a linear combination of basis transformations $V_b^{(l)}\in R^{d^{(l+1)} \times d^{(l)}}$ with coefficients $a_{rb}^{(l)}$ such that only the coefficients depend on $r$.

\par \textbf{Block Diagonal Decomposition}: Block diagonal decomposition introduces a sparse constraint on the weight matrices for each relation type. The core idea is to decompose the latent features into a set of variables that are tightly coupled within groups and loosely coupled between groups. By transforming the large parameter weight matrices into a series of smaller matrices and concatenating them into block diagonal form, block diagonal decomposition ensures the sparsity of the matrix while reducing the number of parameters. Specifically, each weight matrix $W^{(l)}_{r}$ in block diagonal decomposition is defined as the sum of multiple low-dimensional matrices:

\begin{equation}
W_r^{(l)} = \bigoplus_{b=1}^{B} Q_{br}^{(l)}
\end{equation}
where $Q_{br}^{(l)}$ is the submatrix used to construct the block diagonal matrix. This sparse construction effectively reduces the number of parameters and mitigates the risk of overfitting. 
\par Through the RGCN layer, RC-Detection capture the semantic relationships between changed code lines and predict the probability labels indicating whether each deleted code line is the root cause. 

\subsubsection{Deletion Nodes Rankling Layer}
\par Finally, the root cause detection component use the predicted labels obtained from the RGCN layer to rank the deleted line nodes using the RankNet model\cite{Burges_2010}, identifying the deleted lines that are most likely to be the root causes. 
\par The RankNet model is trained to learn the relative priority of the deleted nodes in a pairwise manner. Specifically, for each training pair of nodes $<n_{i},n_{j}>$, RankNet first assigns a score to each node, denoted as $s_{i}$ and $s_{j}$, respectively. Then, the learned probability of $n_{i}$ being ranked higher than $n_{j}$ is calculated as: 
\begin{equation}
P_{ij} = \frac{1}{1+e^{-(s_{i} - s_{j})}}
\end{equation}

Finally, the RankNet model is trained with a Focal loss\cite{lin2017focal} defined as: 
\begin{equation}
L_{focal} = -\alpha_{t}(1-p_{t})^{\gamma} \log(p_{t})
\end{equation}
where $p_{t}$ is the probability of the positive class (i.e., the target category).$\alpha_{t}$ is the weight coefficient for balancing positive and negative samples, addressing the class imbalance issue. $\gamma$ is the adjustment coefficient used to adjust the weight of easy and difficult samples, aimed at reducing the impact of easy to classify samples. $t$ represents the true label. if the sample is positive, then =1; otherwise, =0. 

\par During inference, the trained RankNet model assigns a score to each deleted node, which is directly used to determine the overall priority of the deleted nodes.

\par In this way, RC-Detection complete the detection of root cause deleted lines of bugs in bug-fixing commits. RC-Detection utilize the semantic relationships and node features between each deleted line and other modified code lines to predict their probability of being the root cause, resulting in a ranking for each deleted line. For every deleted line , the higher the ranking, the more likely it is to be the root cause, thus addressing Challenge 2.
 
\section{Experimental Setup}
In this section, we first explore our reasearch questions in Section 4.1. Then, we provided a detailed introduction to the dataset, evaluation metrics, SOTA method, and training details in sections 4.2, 4.3, 4.4, and 4.5, respectively.

\subsection{Research Questions}

Our evaluation explores the following Research Questions (RQs):

\textbf{RQ1:} A comparison of the effectiveness of RC-Detection with other SOTA Methods

\textbf{RQ2:} The impact of different node vectorization methods on RC-Detection

\textbf{RQ3:} Comparing the effectiveness of different graph convolutional layers in RC-Detection

\textbf{RQ4:} Effectiveness comparison of RC-Detection under different parameter settings

\textbf{RQ5:} What is the impact of different imbalanced loss functions on the RC-Detection method?

RQ1 is used to evaluate the superiority of RC-Dtection in identifying root causes of bugs in bug-fixing. 
RQ2 is employed to assess the impact of different node vectorization methods on RC-Detection. 
RQ3 aims to assess the influence of different graph convolutional networks on RC-Detection to its performance. 
RQ4 is used to explore the effect of different parameter settings on RC-Detection. 
RQ5 aims to evaluate the effectiveness of RC-Detection using different imbalanced loss functions.

\subsection{Datasets}
  
In this paper, the datasets are collected  from 87 open-source projects, including 675 bug-fix commits\cite{Wen_Wu_Liu_Tian_Xie_Cheung_Su_2019}, .\cite{Song_Lin_Ng_Wu_Peng_Dong_Mei_2021}, \cite{Neto_Costa_Kulesza_2019}. \tablename~\ref{table1} presents a summary of the overall statistics for the three datasets. In these datasets, each project contains a varying number of bug-fixing commits, ranging from 23 to 222, and the number of bug-inducing commits ranges from 27 to 222. During the data preparation phase, RC-Detection adopted the processing methods used by Tang et al\cite{tang2023neural}. in the experiments to further identify which deleted lines in the bug-fixing commits were the root causes of the bugs and obtained the true labels for these deleted lines.

\subsection{Evaluation Metrics}
To evaluate the performance of identifying the root causes in bug-fixing commits, RC-Detection use the following widely adopted metrics:

\par \textbf{Recall@N}.The Top-N metric indicates the number of bug-fixing commits where at least one root cause deletion line is found within the top N positions of the ranked list. Previous research has shown that developers tend to focus on a small number of elements at the top of the ranked list\cite{Parnin_Orso_2011}. Therefore, RC-Detection consider values of N as 1, 2, and 3. Recall@N is defined as the ratio of the number of correctly predicted bug-fixing commits to the total number of actual bug-fixing commits, under the Top-N constraint. Its formula is expressed as:
\begin{equation}
Recall @ N = \frac{n_{N}}{n_{R}}
\end{equation}
where $n_{R}$ represents the total number of actual bug-fixing commits, which varies depending on the value of N.

\par \textbf{Mean First Rank (MFR)}. For all deleted lines in a commit, the first rank represents the position of the actual top-ranked root cause deletion in the predicted list. MFR calculates the average first rank across all bug-fixing commits. Therefore, the lower the Mean First Rank, the more accurate the prediction results are.


\begin{table}[t]
  \centering
  \caption{An Overview of the Experimental Datasets.}
  \setlength{\tabcolsep}{2mm} 
  \renewcommand{\arraystretch}{1.3}
  \begin{tabular}{ccccc}
    \toprule
    \textbf{Datasets} & \textbf{Projects} & \textbf{Bug-fixing} & \textbf{Bug-inducing}  \\
    \midrule
    \multirow{6}{*}{\textbf{DATASET1}} 
          & Accumulo & 35    & 55  \\
          & Ambari & 38    & 44  \\
          & Hadoop & 53    & 57  \\
          & Lucene & 70    & 145  \\
          & Oozie & 45    & 50  \\
           \cline{2-4}
          & \textbf{Total} & 241   & 351  \\
    \midrule
    \multirow{7}{*}{\textbf{DATASET2}} 
          & Jsoup & 63    & 63  \\
          & Fastjson & 222   & 222  \\
          & Verdict & 53    & 53  \\
          & Closure-templates & 32    & 32  \\
          & Twilio-java & 39    & 39 \\
          & ...(120 more projects) & 548   & 548  \\
          \cline{2-4}
          & \textbf{Total} & 957   & 957  \\
    \midrule
    \multirow{6}{*}{\textbf{DATASET3}} 
          & Mockito & 32    & 53\\
          & Joda-time & 23    & 27  \\
          & Commons-math & 85    & 111  \\
          & Commons-lang & 53    & 65  \\
          & Closure-compiler & 98    & 122  \\
         \cline{2-4}
          & \textbf{Total} & 291   & 378 \\
    \bottomrule
    \end{tabular}%
  \label{table1}%
\end{table}%

\begin{table*}[t]
  \centering
  \caption{(RQ1) The performance comparisons between our method and baselines in ranking deletion lines}
  {\large 
    \begin{tabular}{p{8em}cccc}
    \toprule
    \textbf{Methods} & \multicolumn{1}{p{4.19em}}{\textit{\textbf{Recall@1}}} & \multicolumn{1}{p{4.19em}}{\textit{\textbf{Recall@2}}} & \multicolumn{1}{p{4.19em}}{\textit{\textbf{Recall@3}}} & \multicolumn{1}{p{2em}}{\textit{\textbf{MFR}}} \\
    \midrule
    RF    & 0.694 & 0.811 & 0.882 & 3.295 \\
    LR    & 0.701 & 0.813 & 0.872 & 3.541 \\
    SVM   & 0.714 & 0.806 & 0.869 & 3.215 \\
    XGB   & 0.718 & 0.811 & 0.867 & 3.133 \\
    KNN   & 0.677 & 0.792 & 0.86  & 2.773 \\
    
    Bi-LSTM & 0.656 & 0.746 & 0.82  & 3.448 \\
    
    NEURAL-SZZ & 0.779 & 0.841 & 0.886 & 2.425 \\
    
    RC-Detection & \textbf{0.811} & \textbf{0.884} & \textbf{0.924} & \textbf{1.830} \\
    \bottomrule
    \end{tabular}%
  }
  \label{table2}%
\end{table*}%

\subsection{SOTA Methods}
To validate the effectiveness of our RC-Detection method, we compare it with the Neural SZZ proposed by Tang et al\cite{tang2023neural}. Additionally, we compare our experimental results with the deep learning method Bi-LSTM\cite{Graves_Mohamed_Hinton_2013} and four machine learning methods: RF, LR, SVM, XGB, and KNN. These classifiers compute the probability of each deleted line in a bug-fixing commit being the root cause, and we rank all the deleted lines in the bug-fixing commit based on their probabilities. This ranking is then used to calculate evaluation metrics and compare with each other.

\subsection{Training Details}
\par This section describes the hyperparameters used during the training phase of the RC-Detection method. In the root cause detection component, RC-Detection first use the pre-trained CodeBERT model\cite{Feng_Guo_Tang_Duan_Feng_Gong_Shou_Qin_Liu_Jiang_et_al._2020} from the Hugging Face library to obtain embedding vectors for the graph nodes. Then, we employ two layers of RGCN implemented in PyTorch, with the num\_bases parameter set to 30 in every convolutional layer. Finally, RC-Detection use the RankNet model, also implemented in PyTorch, to rank the nodes, which is the same as\cite{Tan_Zhang_Mi_Cao_Sun_Lin_Yang_2021}. During training, RC-Detection apply 10-fold cross-validation\cite{Fushiki_2011} to evaluate RC-Detection's performance. RC-Detection use Focal Loss\cite{lin2017focal} as the loss function and the Adam optimizer to minimize the loss. RC-Detection is trained for 20 epochs with a batch size of 128. The learning rate is set to 5e-6. For the analysis phase, RC-Detection specify the values of N in the Recall@N metric as 1, 2, and 3.

\section{Experimental Results}

\subsection{RQ1: A comparison of the effectiveness of RC-Detection with other SOTA Methods}
\par \textbf{Motivation.}To validate the effectiveness of our method, we compare the prediction results of RC-Detection with those of Neural SZZ proposed by Tang et al. We also include a deep learning baseline method and four machine learning baseline methods in the comparison to explore whether RC-Detection outperforms these methods.

\par \textbf{Method.} To evaluate the effectiveness of RC-Detection, we conducted experiments using three reliable datasets collected by Wen et al., Song et al., and Neto et al. For a fair comparison, we adopted the same ten-fold cross-validation strategy as proposed by Tang et al\cite{tang2023neural}. Specifically, we randomly shuffled the datasets and then used stratified random sampling to divide them into ten equal-sized folds, ensuring that each fold had a similar distribution of information types. Nine folds were used to train the prediction model (the training dataset), while the remaining fold was used for model evaluation (the test dataset). This entire process was repeated ten times, with each fold serving as the test set once. The average of the ten test results was recorded as the final effectiveness assessment. We then used four metrics to evaluate RC-Detection's prediction results: Recall@1, Recall@2, Recall@3, and Mean First Rank (MFR). A higher Recall@1, Recall@2, and Recall@3, along with a lower MFR, indicates better predictive performance.

\par \textbf{Results.} The effectiveness comparison between RC-Detection and other SOTA methods is presented in \tablename~\ref{table2}, with the best results for each metric highlighted in bold. The experimental results indicate that RC-Detection outperforms all other SOTA methods in Recall@1, Recall@2, Recall@3, and Mean First Rank (MFR).Recall@1 indicates the proportion of bug-fixing commits in which the root cause deletion line ranks first in the predicted list among all actual bug-fixing commits, namely the proportion of completely accurate predictions of the root cause. RC-Detection achieves a Recall@1 scores of 0.811, it means RC-Detection can accurately identify the root causes of bugs in bug-fixing. Recall@2 and Recall@3 represent respectively the proportion of bug-fixing commits in which the root cause deletion line ranks top-2 and top-3 in the predicted list among all actual bug-fixing commits. For this two metric, RC-Detection achieves scores of 0.884 and 0.924. Furthermore, MFR calculates the average first rank across all bug-fixing commits and MFR of RC-Detection is 1.830. 

\par For the machine learning (ML) and deep learning (DL) baselines, the results show that DL methods perform relatively worse compared to ML methods, which aligns with findings by Wu et al. Compared to all DL and ML methods , RC-Detection improved Recall@1 by 12.952\% to 23.628\%, Recall@2 by 8.733\% to 18.499\%, and Recall@3 by 4.762\% to 12.683\%. Additionally, for MFR, the improvements for RC-Detection ranged from 34.006\% to 48.320\%. These notable enhancement in performance robustly substantiates the effectiveness of RC-Detection integrating information from the heterogeneous graph of changed code lines and capturing semantic relationships between changed code lines.

\par For the baselines Tang et al.'s Neural SZZ is the state-of-the-art method for identifying root causes of bugs in bug-fixing. For Recall@1, Recall@2 and Recall@3, RC-Detection improves the performance of Neural SZZ by 4.107\%, 5.113\% and 4.289\%, respectively. Furthermore, RC-Detection outperforms Neural SZZ by 69.8\%, 52.1\%, in MFR. Considering Neural SZZ using heterogeneous graph neural network, the substantial improvement in performance strongly substantiates the effectiveness of our introducing graph type conversion and root cause detection component in RC-Detection. This demonstrates the effectiveness of our method in identifying the root causes of bugs in bug-fixing. 

\par \textbf{Conclusion.} The ten-fold cross-validation experiments conducted on three reliable datasets demonstrate that RC-Detection significantly outperforms other state-of-the-art methods across evaluation metrics. Compared to the best-performing SOTA methods, RC-Detection achieves improvements of 4.107\%, 5.113\%, 4.289\%, and 24.536\% in Recall@1, Recall@2, Recall@3, and MFR, respectively. These enhancements indicate that RC-Detection effectively utilizes the semantic relationships between changed code lines and exhibits superior performance in identifying root causes of bugs in bug-fixing.

\subsection{RQ2: The Impact of Different Node Vectorization Methods on RC-Detection}
\textbf{Motivation.} RC-Detection use CodeBERT to capture the semantic relationships between code statements and embed them into fixed-length vectors. To verify the effectiveness of CodeBERT, we further investigate how different node vectorization methods affect the performance of RC-Detection.

\par \textbf{Method.}To test whether CodeBERT offers an advantage over other node vectorization methods, we compared RC-Detection’s prediction results with three other BERT-based models: ALBERT\cite{lan2019albert}, DistilBERT\cite{sanh2019distilbert}, and RoBERTa\cite{liu2019roberta}. Additionally, we used the standard BERT\cite{devlin2018bert} model as a baseline for comparison. The performance of the models was evaluated using four metrics: Recall@1, Recall@2, Recall@3, and MFR.

\par \textbf{Results.} \tablename~\ref{table3} shows the results of the models using different BERT variants, with the best results for each metric highlighted in bold. The experiments demonstrates that the model utilizing CodeBERT outperformes the other BERT variants across all metrics—Recall@1, Recall@2, Recall@3, and MFR. Specifically, CodeBERT achieves Recall scores of 0.811, 0.884, and 0.924 for top 1, 2, and 3 predictions, respectively, and an MFR score of 1.830. 

For the original standard BERT model, its scores of the metrics Recall@1, Recall@2 and Recall@3 are 0.799, 0.872 and 0.923, respectively. Additionally, it achieves the score of 1.884 in MFR. Compared to the SOTA methods reported in Section 5.1, RC-Detection with BERT model outperforms them by 2.567\% to 21.799\% in Recall@1. For the metric of Recall@2 and Recall@3, RC-Detection with BERT model improves the performance of them by 3.686\% to 16.890\% and 4.176\% to 12.561\%, respectively. For MFR, outperforms them by 22.309\% to 46.795\%. These improvements in performance robustly substantiates the effectiveness of the Node Embedding Layer in the RC-Detection, which indicates that the use of BERT can enhance the effectiveness of identifying root causes of bugs in bug-fixing commit. 

Compared to other node vectorization methods, CodeBERT improved Recall@1 by 1.502\% to 26.128\%, Recall@2 by 1.029\% to 17.241\%, and Recall@3 by 0.108\% to 11.325\%, respectively. Moreover, CodeBERT shows an MFR improvement range of 2.866\% to 36.656\%. These improvements indicate that choosing CodeBERT as the node vectorization method can best improve the overall performance of RC detection among these BERT models.

\par \textbf{Conclusion.} Our RC-Detection method, utilizing CodeBERT for node vectorization, demonstrated a significant advantage over other BERT variants. Therefore, we conclude that the use of CodeBERT improves RC-Detection’s ability to identify the root causes of bugs in bug-fixing commits.

\subsection{RQ3: Comparing the Effectiveness of Different Graph Convolutional Layers in RC-Detection}
\textbf{Motivation.} In the RC-Detection method, we utilize RGCN to capture the semantic relationships between changed code lines. With the continuous evolution and advancement of deep learning models, various versions of graph convolutional neural networks have emerged, and different structures of these networks may affect the predictive performance of RC-Detection. Therefore, we further investigate the effectiveness of the relational graph convolutional network used in this study by comparing it with other graph convolutional network variants.

\par \textbf{Method.} To assess whether our relational graph convolutional network (RGCNConv)\cite{Schlichtkrull_Kipf_Bloem_van·den_Berg_Titov_Welling_2018} holds an advantage over other graph convolutional network variants, we compare RC-Detection's predictions with four additional variants: graph attention network (GATConv)\cite{Liu_Zhou_2020},graph convolutional network (GCNConv)\cite{Schlichtkrull_Kipf_Bloem_van·den_Berg_Titov_Welling_2018}, \cite{10210676}, \cite{feng2023fcgcn}, relational graph attention network(RGATConv)\cite{busbridge2019relational} and generalized graph convolutional network(GENConv)\cite{li2020deepergcn}. Similar to the procedure in Section 5.1, we perform ten-fold cross-validation and evaluate RC-Detection's predictions using the following four metrics: Recall@1, Recall@2, Recall@3, and Mean First Rank (MFR). Finally, we use the relative percentage improvements to compare the results.

\begin{table}[t]
  \centering
  \caption{(RQ2) The performance comparisons of different node vectorization methods in the RC-Detection method.}
  \renewcommand{\arraystretch}{1.3} 
  
    \begin{tabular}{p{6em}cccc}
    \toprule
    \textbf{Vectorization Methods} & \multicolumn{1}{p{4.19em}}{\textit{\textbf{Recall@1}}} & \multicolumn{1}{p{4.19em}}{\textit{\textbf{Recall@2}}} & \multicolumn{1}{p{4.19em}}{\textit{\textbf{Recall@3}}} & \multicolumn{1}{p{2em}}{\textit{\textbf{MFR}}} \\
    \midrule
    BERT  & 0.799 & 0.872 & 0.923 & 1.884 \\
    ALBERT & 0.643 & 0.754 & 0.83  & 2.889 \\
    DistilBERT & 0.787 & 0.875 & 0.905 & 2.05 \\
    RoBERTa & 0.728 & 0.838 & 0.884 & 2.207 \\
    CodeBert & \textbf{0.811} & \textbf{0.884} & \textbf{0.924} & \textbf{1.83} \\
    \bottomrule
    \end{tabular}%
  
  \label{table3}%
\end{table}%

\begin{table}[t]
  \centering
  \caption{(RQ3)The performance comparisons of different graph convolutional layers in the RC-Detection method.}
   \renewcommand{\arraystretch}{1.3} 
   
    \begin{tabular}{p{6em}cccc}
    \toprule
    \textbf{Convolutional Layer} & \multicolumn{1}{p{4.19em}}{\textit{\textbf{Recall@1}}} & \multicolumn{1}{p{4.19em}}{\textit{\textbf{Recall@2}}} & \multicolumn{1}{p{4.19em}}{\textit{\textbf{Recall@3}}} & \multicolumn{1}{p{2em}}{\textit{\textbf{MFR}}} \\
    \midrule
    GAT   & 0.801 & 0.882 & 0.921 & 1.922 \\
    GCN   & 0.784 & 0.874 & 0.911 & 1.867 \\
    GEN   & 0.79  & 0.866 & 0.912 & \textbf{1.828} \\
    RGAT  & 0.787 & 0.872 & 0.914 & 1.949 \\
    RGCN  & \textbf{0.811} & \textbf{0.884} & \textbf{0.924} & 1.83 \\
    \bottomrule
    \end{tabular}%
 
  \label{table4}%
\end{table}%

\par \textbf{Results.} \tablename~\ref{table4} presents the results and performance comparisons of models using different graph convolutional layers, with the best results for each metric highlighted in bold. The experimental results indicate that the RC-Detection method utilizing the RGCNConv achieves the best performance in Recall@1, Recall@2, and Recall@3, surpassing models that employ other graph convolutional layers. Specifically, it achieves Recall scores of 0.811, 0.884, and 0.924 for the top 1, 2, and 3, respectively. Additionally, RC-Detection with RGCNConv obtains an MFR score of 1.830, which is very close to the best score of 1.828. 

\par Compared to SOTA methods using other graph convolutional layers, RC-Detection improves Recall@1 by 1.248\% to 3.444\%, Recall@2 by 0.227\% to 2.079\%, and Recall@3 by 0.326\% to 1.427\%. Although RC-Detection does not achieve the absolute best performance for MFR, the difference from the highest model score is minimal, and its improvement over several other SOTA methods ranges from 1.982\% to 6.106\%. These results demonstrate that the RGCNConv used in RC-Detection significantly outperforms other convolutional layers in capturing semantic relationships between changed code lines. 

\par For Recall@1, the scores of the five different graph convolutional layers are 0.801, 0.784, 0.790, 0.787, and 0.811, all of which exceed the performance of other SOTA methods reported in Section 5.1. This indicates that the use of graph convolutional networks better captures the semantic relationships between changed code lines, thereby enhancing the effectiveness of identifying root causes in bug-fixing submissions.

\par \textbf{Conclusion.} Our RC-Detection method, which utilizes relational graph convolutional network(RGCNConv), demonstrates a significant advantage in comparison experiments with other graph convolutional layers. Furthermore, the experiments confirm that graph neural networks are effective in capturing the semantic relationships between changed code lines, leading to improved detection of root cause deletion lines in bug-fixing submissions.

\subsection{RQ4: Effectiveness Comparison of RC-Detection under Different Parameter Settings}
\textbf{Motivation.} In Section 5.3, we determined that RC-Detection using relational graph convolutional network (RGCNConv) by comparing different graph convolutional networks, has a significant advantage over other graph convolutional layers. However, the number of layers in the RGCN can also impact model performance. Additionally, to address the overfitting problem caused by the explosion of model parameters as the number of relations increases, RGCN introduces two regularization methods (basis decomposition and block diagonal decomposition) that cannot be applied within the same convolutional layer. Therefore, this section focuses on investigating two aspects of RC-Detection's parameter settings: the number of layers in the relational graph convolutional network and the effects of different configurations of its unique regularization methods on the prediction performance of RC-Detection.

\par \textbf{Method.} To explore the optimal number of layers for the relational graph convolutional layer used in RC-Detection, we compared models with layer counts ranging from 1 to 5: $\text{RC-Detection}_{layer=1}$, $\text{RC-Detection}_{layer=2}$, $\text{RC-Detection}_{layer=3}$, $\text{RC-Detection}_{layer=4}$, and RC-Detection$_{layer=5}$. Additionally, after determining the number of layers in the relational graph convolutional layer, we conducted experiments with different settings for the two regularization methods (basis decomposition and block diagonal decomposition) to identify the configuration that yields the best model performance. We evaluated the models using four metrics: Recall@1, Recall@2, Recall@3, and Mean First Rank (MFR). Our primary focus was on the Recall@1 score, as it indicates the model's performance in accurately predicting the root causes of bugs.

\par \textbf{Results.} \tablename~\ref{table5} presents the experimental results regarding the number of layers in the relational graph convolutional layer. The results indicate that $\text{RC-Detection}_{layer=2}$ achieved the best performance on Recall@1 and Recall@2, with scores of 0.811 and 0.924, respectively. Meanwhile, $\text{RC-Detection}_{layer=1}$ demonstrated the highest performance on Recall@2 and MFR, with scores of 0.887 and 1.816. We observed a trend of decreasing performance across all metrics as the number of relational graph convolutional layers increased beyond two. However, for the Recall@1 metric, the $\text{RC-Detection}_{layer=2}$ model outperformed other models by 1.122\%, 1.502\%, 4.242\%, and 4.645\%. This demonstrates that $\text{RC-Detection}_{layer=2}$ (RC-Detection with two relational graph convolutional layers) is particularly effective in identifying the root causes of bugs in bug-fixing. Therefore, we proceeded to explore different settings for the two regularization methods based on having set the number of relational graph convolutional layers to 2.

\begin{table}[t]
  \centering
  \caption{(RQ4) The performance comparisons of different numbers of convolutional layers in the RC-Detection method.}
  \renewcommand{\arraystretch}{1.3}
    \begin{tabular}{p{10em}cccc}
    \toprule
    \textbf{Models} & \multicolumn{1}{p{4.19em}}{\textit{\textbf{Recall@1}}} & \multicolumn{1}{p{4.19em}}{\textit{\textbf{Recall@2}}} & \multicolumn{1}{p{4.19em}}{\textit{\textbf{Recall@3}}} & \multicolumn{1}{p{2em}}{\textit{\textbf{MFR}}} \\
    \midrule
    $\text{RC-Detection}_{layer=1}$ & 0.802 & \textbf{0.887} & 0.921 & \textbf{1.816} \\
    $\text{RC-Detection}_{layer=2}$ & \textbf{0.811} & 0.884 & \textbf{0.924} & 1.830 \\
    $\text{RC-Detection}_{layer=3}$ & 0.799 & 0.874 & 0.912 & 1.890 \\
    $\text{RC-Detection}_{layer=4}$ & 0.778 & 0.859 & 0.911 & 2.085 \\
    $\text{RC-Detection}_{layer=5}$ & 0.775 & 0.850  & 0.899 & 2.124 \\
    \bottomrule
    \end{tabular}%
  \label{table5}%
\end{table}%

\begin{table}[t]
  \centering
  \caption{(RQ4) The performance comparisons of different regularization method settings in the RC-Detection method.( The basis represents basis- decomposition and the block represents block-diagonal-decomposition)}
  \renewcommand{\arraystretch}{1.3}
    \begin{tabular}{p{11em}cccc}
    \toprule
    \textbf{Models} & \multicolumn{1}{p{4em}}{\textit{\textbf{Recall@1}}} & \multicolumn{1}{p{4em}}{\textit{\textbf{Recall@2}}} & \multicolumn{1}{p{4em}}{\textit{\textbf{Recall@3}}} & \multicolumn{1}{p{2em}}{\textit{\textbf{MFR}}} \\
    \midrule
    RC-Detection$_{basis\&basis}$ & \textbf{0.811} & 0.884 & 0.924 & 1.830 \\
    RC-Detection$_{basis\&block}$ & 0.790  & 0.881 & 0.920  & 1.907 \\
    RC-Detection$_{block\&basis}$ & 0.805 & \textbf{0.890} & \textbf{0.927} & 1.816 \\
    RC-Detection$_{block\&block}$ & 0.805 & \textbf{0.890} & \textbf{0.927} & \textbf{1.793} \\
    \bottomrule
    \end{tabular}%
  \label{table6}%
\end{table}%

\par Having established that the number of layers in the relational graph convolutional network is 2, we denoted $basis$ for basis decomposition and $block$ for block diagonal decomposition to train four different models: RC-Detection$_{basis\&basis}$, RC-Detection-$_{basis\&block}$, RC-Detection-$_{block\&basis}$, and RC-Detection-$_{block\&block}$. \tablename~\ref{table6} displays the results and performance comparisons of RC-Detection using different combinations of the basis decomposition and block diagonal decomposition. The experimental results show that RC-Detection$_{basis\&basis}$ achieved the best performance on Recall@1, with a score of 0.811. Additionally, RC-Detection-$_{block\&block}$ and RC-Detection-$_{block\&basis}$ were tied for first place on Recall@2 and Recall@3, with scores of 0.890 and 0.927, respectively. The model with the highest MFR score was RC-Detection-$_{block\&block}$, with a score of 1.793. Comparing the results of the various models across Recall@2, Recall@3 and MFR, we found that, except for RC-Detection-$_{basis\&block}$, which performed lower than the others, the performance differences among the other models were relatively small. However, on the crucial Recall@1 metric, RC-Detection$_{basis\&basis}$ outperformed the other models by 2.658\%, 0.745\%, and 0.745\%. Thus, we conclude that using basis function decomposition as a regularization method in both the first and second layers is more suitable for accurately predicting the root causes of bugs.

 \par \textbf{Conclusion.} Our RC-Detection method, utilizing two layers of graph convolution, demonstrated significant advantages in comparative experiments with other layer configurations. The experiments revealed that employing basis function decomposition in both convolutional layers effectively mitigates the risk of overfitting caused by the increased number of parameters with the growing number of relations, leading to optimal performance in detecting root cause deletion lines of bugs in bug-fixing commits.

 \subsection{RQ5: What is the impact of different imbalanced loss functions on the RC-Detection method?}
\textbf{Motivation.} The dataset used in our experiments contains class imbalance, where the number of root cause lines that introduce bugs in bug-fixing commits is much lower than other code lines. Additionally, the dataset includes samples of root cause lines that are either easy or difficult to distinguish. This distribution may prevent RC-Detection from adequately learning the features of each class, thus affecting the prediction performance. To address this issue, RC-Detection employs the Focal loss function. In this section, we test various loss functions to compare their impact on model effectiveness.

\par \textbf{Method.}We tested several commonly used loss functions to handle the issue of class imbalance: weighted binary cross-entropy loss\cite{rezaei2020addressing}, Gradient Harmonizing Mechanism loss\cite{li2019gradient}, and Focal loss\cite{lin2017focal}. Additionally, we used unweighted binary cross-entropy loss\cite{ruby2020binary} as a baseline for comparison. To evaluate the impact of different loss functions, we analyzed the experimental results using four evaluation metrics: Recall@1, Recall@2, Recall@3, and MFR. Among these, we focused primarily on the Recall@1 score, as it represents RC-Detection's ability to accurately predict the root cause of bugs.

\par \textbf{Results.} \tablename~\ref{table7} shows the results and performance comparisons of RC-Detection using different imbalance loss functions, with the best result for each metric highlighted in bold. Among the evaluated loss functions, the RC-Detection trained with Focal loss exhibited the best performance across the Recall@1, Recall@2, and Recall@3 metrics, scoring 0.811, 0.884, and 0.924, respectively. Additionally, RC-Detection using Focal loss achieved a score of 1.830 on the MFR metric, ranking second but only 0.328\% behind the best score of 1.824. However, in the most critical metric, Recall@1, RC-Detection trained with Focal loss improved by 1.629\%, 1.502\%, and 1.629\% compared to other loss functions, respectively. This demonstrates the superior effectiveness of RC-Detection trained with Focal loss in identifying the root causes of bugs in bug-fixing.

\par For BCE With Weight loss, it achieved the best score in the MFR metric and tied for first in Recall@2, while ranking second in Recall@1 and Recall@3. Since MFR represents the average rank of the actual top-ranked root-cause deletion lines in the prediction list, a lower value indicates better predictive performance. Therefore, the model trained with BCE With Weight loss also demonstrated effectiveness in identifying the root causes of errors in bug-fixing. However, as its performance in the most critical Recall@1 metric was similar to the baseline BCE and both were lower than Focal loss, we conclude that Focal loss is more effective in mitigating the impact of imbalance data distribution on the results. Thus, compared to other loss functions, Focal loss is better suited for root cause detection.

\par \textbf{Conclusion.}The Focal loss function used in RC-Detection significantly outperformed other loss functions in Recall@1. Furthermore, models trained with this loss function were able to more accurately predict the root causes of most errors in bug-fixing. Therefore, the Focal loss function effectively alleviated the issue of uneven difficulty in distinguishing the root causes of bugs within varying code data.

\begin{table}[t]
  \centering
  \caption{(RQ5) The performance comparisons of different imbalance loss functions in the RC-Detection method.}
  \renewcommand{\arraystretch}{1.3}
    \begin{tabular}{p{6em}cccc}
    \toprule
    \textbf{Loss} & \multicolumn{1}{p{4.19em}}{\textit{\textbf{Recall@1}}} & \multicolumn{1}{p{4.19em}}{\textit{\textbf{Recall@2}}} & \multicolumn{1}{p{4.19em}}{\textit{\textbf{Recall@3}}} & \multicolumn{1}{p{4.19em}}{\textit{\textbf{MFR}}} \\
    \midrule
    BCE   & 0.798 & 0.869 & 0.915 & 1.89 \\
    BCEWithWeight & 0.799 & \textbf{0.884} & 0.923 & \textbf{1.824} \\
    GHM   & 0.789 & 0.875 & 0.915 & 1.962 \\
    Focal & \textbf{0.811} & \textbf{0.884} & \textbf{0.924} & 1.83 \\
    \bottomrule
    \end{tabular}%
    \label{table7}%
\end{table}%

\section{Threats to Validity}
In this section, we analyze the threats to validity from three perspectives: internal validity, external validity, and construct validity.

\subsection{Internal validity}
The main threat to internal validity involves the accuracy of reproducing state-of-the-art methods. Ensuring accurate replication of these models is crucial for internal validity. To reproduce SOTA methods, we used the authors' provided code and attempted to replicate the experimental setup as accurately as possible to ensure reliable reproduction of the results.
\par Another potential threat to internal validity is the accuracy and reproducibility of the RC-Detection model's performance. To address this, we thoroughly reviewed our code, with a particular focus on the root cause detection component. Additionally, we plan to open-source our code so that other researchers can replicate and validate our study.

\subsection{External Validity}
The main threat to external validity in this work comes from the limitations of the datasets. Although we combined three reliable datasets collected by Wen et al., Song et al., and Neto et al., the total number of bugs in the dataset is limited, with only 675 commits in total. Another potential threat is that our study includes only Java projects. To improve external validity and enhance the model's generalization ability, future work will involve validating the model on broader datasets and across projects in various programming languages.

\subsection{Construct Validity}
The potential threat to construct validity relates to the evaluation metrics used in the RC-Detection model. To mitigate this, we used Recall@N and Mean First Rank (MFR) to evaluate the model's effectiveness, metrics commonly employed for ranking algorithms, which reduces the threat to construct validity in our work.	 

\section{Related Work}

\par In this section, we summarize the research relevant to our study.

\subsection{SZZ algorithm}
In the field of software engineering, Just-In-Time (JIT) defect prediction models are employed to identify potential defects in code commits in real-time. The goal is to automatically defect risky changes before code reviews, thereby minimizing manual inspection effort and improving software quality. JIT defect prediction model typically involves three key stages: data labeling, feature extraction, and model training. The data labeling process commonly utilizes the SZZ algorithm, which integrates version control systems (VCS) and issue tracking systems (ITS) to trace back and identify bug-inducing changes. The original SZZ algorithm (B-SZZ), introduced by Sliwerski et al.\cite{sliwerski2005changes}, traces bug-fixing commits to identify which lines, either deleted or modified, introduced the defects, marking them as bug-inducing commits.

\par However, the precision of the SZZ algorithm is often affected by noise, leading to misclassifications. To address this, several enhanced versions of SZZ have been proposed through static analysis techniques aimed at noise reduction. For example, Kim et al. introduced AG-SZZ\cite{Kim_Zimmermann_Pan_Jr_Whitehead_2006}, which improves upon B-SZZ by filtering out non-functional changes such as blank lines, comment lines, and formatting changes using annotation graphs. Similarly, Da Costa et al. developed MA-SZZ\cite{da_Costa_McIntosh_Shang_Kulesza_Coelho_Hassan_2017}, focusing on filtering meta-changes from bug-inducing commits. Neto et al\cite{Neto_da_Costa_Kulesza_2018}. integrated refactoring detection tools, such as RefDiff and Refactoring Miner, into the RA-SZZ algorithm to enhance its ability to filter out false positives due to refactoring.

\par Different from these traditional methods, we propose a model based on Relational Graph Convolutional Networks (RGCN), which captures the semantic relationships between changed code lines. This method allows us to detect the root cause of potential bugs in deletion lines, effectively reducing noise in the SZZ algorithm.

\subsection{Bug-fixing commit and its related applications}
A bug-fixing commit refers to a code submission made during the software development process to correct defects. It documents the detailed process of fixing bugs and plays a significant role in areas such as defect prediction and software analysis. Numerous studies rely on bug-fixing commits for various purposes, including bug localization, automated repair, commit history analysis, and automated defect prediction. 

\par For example, Wong et al\cite{wong2010regression}. introduced a regression test selection method to automatically identify the lines of code associated with bug-fixing commits, helping developers reduce the debugging scope. Additionally, automated defect prediction models based on historical bug-fixing commits have become a primary research direction in this domain. Kamei et al. proposed the Just-In-Time (JIT)\cite{Kamei_Shihab_Adams_Hassan_Mockus_Sinha_Ubayashi_2013} defect prediction model, which analyzes the features of historical commits—such as code characteristics and developer behavior—to predict potential defects in new commits in real time. Furthermore, automated bug repair has emerged as another significant area of bug-fixing commit research. One notable example is GenProg\cite{le2011genprog}, introduced by Weimer et al. GenProg is an automated program repair technique based on genetic programming that uses historical bug-fixing commits to generate repair patches and validates the correctness of these patches through testing.

\section{Conclusions and Future Work}
In this paper, we proposed RC-Detection, a method based on Graph Neural Networks, to capture the semantic relationships between changed code lines for predicting the root causes of errors in bug-fixing commits. RC-Detection mainly consists of three components: the graph construction component, the graph type conversion component, and the root cause detection component. The graph construction component builds a heterogeneous graph by analyzing the source code of bug-fixing commits. The graph type conversion component unifies heterogeneous graph data into homogeneous graph data, facilitating the integration of information from different nodes and edges. The root cause detection component uses a relational graph convolutional network to capture the semantic relationships between changed code lines and ultimately identifies the root-cause deletion lines in bug-fixing commits by ranking the deleted code lines. Experimental results show that our RC-Detection method outperforms state-of-the-art methods across four evaluation metrics (Recall@1, Recall@2, Recall@3, and MFR). Compared to other SOTA methods, RC-Detection improved Recall@1 by 4.107\% to 23.628\%, Recall@2 by 5.113\% to 18.499\%, and Recall@3 by 4.289\% to 12.683\%. Additionally, for MFR, RC-Detection achieved an improvement range from 24.536\% to 48.320\%. These improvements demonstrate the effectiveness of RC-Detection in identifying the root causes of bugs in bug-fixing commits.

\par In future work, we plan to explore different graph construction methods and various Graph Neural Network models with the aim of capturing richer and more accurate semantic relationships between changed code lines to improve model performance. Additionally, we intend to expand our research by collecting more high-quality datasets and training our model on projects in various programming languages.

\section*{Acknowledgment}

This work was supported by the National Natural Science Foundation of China (No.62472062, No.62202079), the Dalian Excellent Young Project (2022RY35), the Fundamental Research Funds for the Central Universities (No.3132024257).

\bibliographystyle{unsrt}
\bibliography{reference}
\end{document}